\documentclass[twocolumn]{aa} 
\usepackage{ulem} 
\usepackage{natbib,twoopt}
\usepackage[varg]{txfonts} 
\bibpunct{(}{)}{;}{a}{}{,}
\usepackage{booktabs} 

\usepackage{graphicx}
\usepackage{float}
\usepackage{lscape}
\usepackage{pdflscape}
\usepackage{sidecap}
\usepackage{xcolor}
\usepackage{comment}
\usepackage{rotating}
\usepackage{siunitx}
\usepackage{soul}
\usepackage{cancel}

\usepackage[utf8]{inputenc}

\usepackage{academicons}
\definecolor{orcidlogocol}{HTML}{A6CE39}
\usepackage{placeins}
\usepackage{subfigure} 
\usepackage{hyperref}

\begin{document}
\title{\texttt{VAR-PZ}: Constraining the photometric redshifts of quasars using variability}
\titlerunning{Constraining the Photometric Redshifts of Quasars using Variability}
\author{
Sarath Satheesh-Sheeba \inst{1}\thanks{E-mail: 1998sarath.ss@gmail.com}~\href{https://orcid.org/0009-0003-0654-6805}{\includegraphics[scale=0.05]{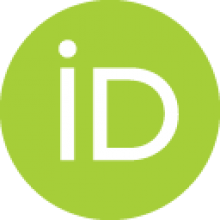}} \and
Roberto J. Assef\inst{2}~\href{https://orcid.org/0000-0002-9508-3667}{\includegraphics[scale=0.05]{Figures/orcid-ID.png}} \and
Timo Anguita\inst{1,3}~\href{https://orcid.org/0000-0003-0930-5815}{\includegraphics[scale=0.05]{Figures/orcid-ID.png}} \and
Paula Sánchez-Sáez\inst{4} ~\href{https://orcid.org/0000-0003-0820-4692}{\includegraphics[scale=0.05]{Figures/orcid-ID.png}}  \and
Raphael Shirley\inst{5} ~\href{https://orcid.org/0000-0002-1114-0135}{\includegraphics[scale=0.05]{Figures/orcid-ID.png}}  \and
Tonima T. Ananna \inst{6} ~\href{https://orcid.org/0000-0001-8211-3807}{\includegraphics[scale=0.05]{Figures/orcid-ID.png}} \and
Franz E. Bauer \inst{7}~\href{https://orcid.org/0000-0002-8686-8737}{\includegraphics[scale=0.05]{Figures/orcid-ID.png}} \and
Alexey Bobrick \inst{8,9}~\href{https://orcid.org/0000-0002-4674-0704}{\includegraphics[scale=0.05]{Figures/orcid-ID.png}}  \and
Carlos Guillermo Bornancini \inst{10,11}~\href{https://orcid.org/0000-0001-6800-3329}{\includegraphics[scale=0.05]{Figures/orcid-ID.png}}\and
Sarah E.~I.~Bosman\inst{12,13} ~\href{https://orcid.org/0000-0001-8582-7012}{\includegraphics[scale=0.05]{Figures/orcid-ID.png}} \and
W.N. Brandt \inst{14,15,16} ~\href{https://orcid.org/0000-0002-0167-2453}{\includegraphics[scale=0.05]{Figures/orcid-ID.png}} \and
Demetra De Cicco\inst{17,3,18}~\href{https://orcid.org/0000-0001-7208-5101}{\includegraphics[scale=0.05]{Figures/orcid-ID.png}} \and
Bozena Czerny \inst{19} ~\href{https://orcid.org/0000-0001-5848-4333}{\includegraphics[scale=0.05]{Figures/orcid-ID.png}}\and
Marta Fatovi\'c\inst{20, 21,18}~\href{https://orcid.org/0000-0003-1911-4326}{\includegraphics[scale=0.05]{Figures/orcid-ID.png}} \and
Kohei Ichikawa \inst{22,23}~\href{https://orcid.org/0000-0002-4377-903X}{\includegraphics[scale=0.05]{Figures/orcid-ID.png}} \and
Dragana Ili\'c\inst{24,25}~\href{https://orcid.org/0000-0002-1134-4015}{\includegraphics[scale=0.05]{Figures/orcid-ID.png}} \and
Andjelka B. Kova\v cevi\'c \inst{24}~\href{https://orcid.org/0000-0001-5139-1978}{\includegraphics[scale=0.05]{Figures/orcid-ID.png}}\and
Guodong Li \inst{26}~\href{https://orcid.org/0000-0003-4007-5771}{\includegraphics[scale=0.05]{Figures/orcid-ID.png}} \and
Mai Liao \inst{2}~\href{https://orcid.org/0000-0002-9137-7019}{\includegraphics[scale=0.05]{Figures/orcid-ID.png}} \and
Alejandra Rojas-Lilayú \inst{27}~\href{https://orcid.org/0000-0003-0006-8681}{\includegraphics[scale=0.05]{Figures/orcid-ID.png}}  \and
Marcin Marculewicz \inst{6} ~\href{https://orcid.org/0000-0002-1380-1785}{\includegraphics[scale=0.05]{Figures/orcid-ID.png}} \and
Daniel Marsango\inst{28}~\href{https://orcid.org/0000-0001-5465-0824}{\includegraphics[scale=0.05]{Figures/orcid-ID.png}} \and
Chiara Mazzucchelli\inst{2}~\href{https://orcid.org/0000-0002-5941-5214}{\includegraphics[scale=0.05]{Figures/orcid-ID.png}} \and
Tatevik Mkrtchyan \inst{2} ~\href{https://orcid.org/0009-0000-6071-4353}{\includegraphics[scale=0.05]{Figures/orcid-ID.png}} \and
Swayamtrupta Panda \inst{29} ~\href{https://orcid.org/0000-0002-5854-7426}{\includegraphics[scale=0.05]{Figures/orcid-ID.png}} \and
Alessandro Peca \inst{30,31}~\href{https://orcid.org/0000-0003-2196-3298}{\includegraphics[scale=0.05]{Figures/orcid-ID.png}}  \and
Bindu Rani \inst{32,33}~\href{https://orcid.org/0000-0001-5711-084X}{\includegraphics[scale=0.05]{Figures/orcid-ID.png}}\and
Claudio Ricci \inst{34,2}~\href{https://orcid.org/0000-0001-5231-2645}{\includegraphics[scale=0.05]{Figures/orcid-ID.png}}\and
Gordon T. Richards \inst{35}~\href{https://orcid.org/0000-0002-1061-1804}{\includegraphics[scale=0.05]{Figures/orcid-ID.png}} \and
Mara Salvato \inst{5,36} ~\href{https://orcid.org/0000-0001-7116-9303}{\includegraphics[scale=0.05]{Figures/orcid-ID.png}} \and
Donald P. Schneider \inst{14,15}~\href{https://orcid.org/0000-0001-7240-7449}{\includegraphics[scale=0.05]{Figures/orcid-ID.png}} \and
Matthew J. Temple \inst{37} ~\href{https://orcid.org/0000-0001-8433-550X}{\includegraphics[scale=0.05]{Figures/orcid-ID.png}} \and
Francesco Tombesi \inst{38,39,40}~\href{https://orcid.org/0000-0002-6562-8654}{\includegraphics[scale=0.05]{Figures/orcid-ID.png}} \and
Weixiang Yu \inst{41}~\href{https://orcid.org/0000-0003-1262-2897}{\includegraphics[scale=0.05]{Figures/orcid-ID.png}} \and
Ilsang Yoon \inst{42,43} ~\href{https://orcid.org/0000-0001-9163-0064}{\includegraphics[scale=0.05]{Figures/orcid-ID.png}}\and
Fan Zou \inst{44}~\href{https://orcid.org/0000-0002-4436-6923}{\includegraphics[scale=0.05]{Figures/orcid-ID.png}} 
}

\institute{\centering \textit{(Affiliations can be found after the references)}}
\authorrunning{Satheesh-Sheeba et al}
\date{Accepted XXX. Received YYY; in original form ZZZ}
\abstract{
 The Vera C. Rubin Observatory's Legacy Survey of Space and Time (LSST) is expected to obtain observations of over ten million quasars. The survey's exceptional cadence and sensitivity will enable a significant fraction of these objects to be monitored in the $ugrizy$ bands, spanning observed wavelengths of approximately $0.3-1.0\,\mu\mathrm{m}$. The unprecedented number of sources  makes spectroscopic follow-up for the vast majority of them unfeasible in the near future, so most studies will have to rely on photometric redshift estimates which are traditionally much less 
 reliable for Active Galactic Nuclei (AGNs) than for inactive galaxies. This work presents a novel methodology to constrain the  photometric redshift of AGNs that leverages the effects of cosmological time dilation, and of the luminosity and wavelength dependence of AGN variability. Specifically, we assume that the variability can be modeled as a damped random walk (DRW) process, and we adopted a parametric model to characterize the DRW timescale ($\tau$) and asymptotic amplitude of the variability (SF{$_\infty$}) based on the redshift, the rest-frame wavelength, and the AGN luminosity. We constructed variability-based photometric redshift (photo-$z$) priors by modeling the observed variability using the expected DRW parameters at a given redshift. These variability-based photo-$z$ (\texttt{VAR-PZ}) priors were then combined with traditional spectral energy distribution (SED) fitting to improve the redshift estimates from SED fitting.

Validation was performed using observational data from the Sloan Digital Sky Survey (SDSS), demonstrating significant reduction in catastrophic outliers by more than 10\% in comparison with SED fitting techniques and improvements in redshift precision. The simulated light curves with both SDSS and LSST-like cadences and baselines confirm that \texttt{VAR-PZ} will be able to constrain the photometric redshifts of SDSS-like AGNs by bringing the outlier fractions down to below 15\% from 32\% (SED alone) at the end of the survey. 
}


\maketitle

\section{Introduction}
Active galactic nuclei (AGNs) are among the most luminous and distant objects in the universe. They are detected in wide redshift ranges, serving as light houses that can be observed over vast cosmic distances, providing crucial knowledge about the early universe \citep[e.g,][]{Richards2006,Fan2023}. The intense luminosity of these objects arises from the accretion of matter onto the central supermassive black hole (SMBH), which is thought to exist in almost every massive galaxy. This accretion process drives the emission of intense radiation across the entire electromagnetic spectrum \citep[e.g, see the review by][]{Padovani2017}. 

Active galactic nuclei (AGNs) are intrinsically variable sources, displaying fluctuations in brightness. These variations in brightness are stochastic, occurring on timescales varying from days to hundreds of years \citep{Vandenberk2004,Sartori2018}. Their flux variability can be used as a tool to probe the central engine’s structure and the physical processes in its immediate environment, as reviewed in detail by \citet{Cackett_2021}. Various previous studies have shown that at least a portion of the observed variations originates directly from the accretion disk itself, rather than being solely a reprocessed response to the highly variable X-ray emission for both long and short timescales \citep[e.g,][]{arevalo2023universalpowerspectrumquasars}. Several studies have made significant progress in linking the variability parameters to basic physical properties through the application of complex methods on large samples of longer and densely sampled light curves \citep[e.g,][]{Vandenberk2004, MacLeod_2010, Simm2016, SanchezSaez2018,Li2018,Luo2020, Tachibana2020,Burke2021,Tang_2023,DHO_Yu_2025}. 

 In order to place AGNs in a cosmological time frame and to study their intrinsic luminosities, black hole masses, accretion rates and their evolution, accurate redshift measurements are necessary. Accurate redshifts are also important for studying AGN feedback and its regulatory role in star formation \citep{Silk_Rees,Fabian_2012}, the coevolution of supermassive black holes and their host galaxies \citep{Kormendy2013}, and their contribution to the ionizing background during the epoch of reionization \citep{Madau_2015}. While current and upcoming multi-object spectroscopic surveys such as the Sloan Digital Sky Survey (SDSS)-V (\citealt{York2000,Kollmeier_2026}), the 4-meter Multi-Object Spectroscopic Telescope (4MOST; \citealt{DeJong2019}), the Dark Energy Spectroscopic Instrument (DESI; \citealt{DESI2016}), the Prime Focus Spectrograph (PFS; \citealt{Tamura_2016}), and the Multi Object Optical and Near-infrared Spectrograph (MOONS; \citealt{Cirasuolo_2012}) will significantly increase the number of observed sources with accurate redshifts, the vast majority of AGNs detected in multiwavelength all-sky and large surveys, such as WISE \citep{Wright2010} and eROSITA (\citealt{Merloni2012,Predehl2021}), will remain without spectroscopic follow-up \citep{Dahlen2013}. This difference will only increase with the current deep and wide-area surveys such as Euclid \citep{Euclid} and the Vera C. Rubin Observatory Legacy Survey of Space and Time (LSST; \citealt{Ivezic2019}) covering unprecedented numbers (\citealt{Newman2022,Savic2023,Li_2025}). Consequently, for most AGNs, photometric redshifts (photo-$z$) remain the only viable alternative to estimate their distances.

Spectral energy distribution (SED) fitting is a widely used method for estimating photo-$z$s for AGNs (e.g, \citealt{Arnouts1999,Ilbert2006}), as implemented in \texttt{LePHARE}, 
which has been optimized for AGNs (e.g, \citealt{Shirley_submitted,2008EASY,salvato2011dissecting,salvato2019many,Salvato2022}). However, factors such as non simultaneous observations, intrinsic variability and mostly featureless SED continuum of the luminous type 1 AGNs affect the redshift estimation based on their broad spectral characteristics. However, the power-law continuum SED of luminous type 1 AGNs lacks the strong spectral features needed to accurately estimate the redshifts, with the notable exception of the Lyman break and Intergalactic medium (IGM) absorption features short of Ly$\alpha$. Combined with the intrinsic variability of these objects, obtaining accurate photo-$z$s from SED modeling is very challenging. Arguably, large host galaxy fractions should make the task much easier, as they would add those missing features \citep{Hsu2014,Temple2021}), so the problem is most critical for luminous quasars. These issues result in many sources exhibiting very broad redshift probability distribution functions (PDFs) that are sometimes multi-peaked, leading to multiple possible redshift solutions. As reported by \citet{Saxena_2024}, SED fitting requires many photometric data points across a wide wavelength range to resolve these degeneracies \citep[e.g,][]{Brescia2019}. SED fitting is able to perform well when it is combined with appropriate redshift priors, such as redshift-dependent absolute magnitude constraints and carefully selected spectral templates \citep{ Fotopoulou2012, Hsu2014, Ananna2017,Peca2021}. Several studies have demonstrated that photometry from narrow and intermediate bands is more effective in this context \citep{salvato2009photometric, Cardamone2010, Hsu2014,JPAS,SPLUS2022}. These bands are more sensitive to emission-line features, allowing for improved identification of line intensities, thereby distinguishing star-forming sources from AGNs. In addition to traditional SED fitting, machine learning (ML) has emerged as a powerful alternative for redshift estimation (e.g, \citealt{Ananna2017, Saxena_2024,Roster2024}). ML models can be sensitive to more subtle features than SED modeling, resulting in higher accuracy for sources that are well represented within the training dataset. However, they also tend to produce a significantly higher fraction of outliers when applied to sources outside the training sample’s parameter space \citep{Duncan2018, Salvato2022}. As these methods often rely on deep and homogenized photometry with deep spectroscopy, they are typically most applicable in pencil-beam surveys. 

In this work, we present \texttt{VAR-PZ}, a variability-based method to constrain the photometric redshifts of AGNs. It has been developed to help overcome the degeneracies of photo-$z$ solutions, reducing  the fraction of outliers and improve the accuracy of the redshifts. It models the AGN variability, including its dependence on luminosity, rest-frame wavelength, and the effects of cosmological time dilation. Assuming the variability follows a DRW process, this work adopts a parametric model from \citet[][hereafter M10]{MacLeod_2010}, where the DRW timescale and amplitude are functions of the redshift, the AGN luminosity, and the rest-frame wavelength. SED decomposition is used to estimate the intrinsic AGN luminosity and account for host galaxy contamination. These parameters are used to construct redshift-dependent priors.
\texttt{VAR-PZ} generates PDFs, which are complementary to those derived from other established techniques and can be integrated with them, such as SED fitting and ML-based methods to improve redshift predictions. In particular, we explore the impact of using the \texttt{VAR-PZ} priors in two key quantities of photo-$z$ estimates: the fraction of outliers, and the accuracy for the non-outliers.
 
The paper is structured as follows.
Section \ref{data} describes the data used to develop and test our algorithm. Section \ref{Methodology} gives an overview of our methodology and the way it is structured. A description of our code is discussed in Section \ref{Code}. Section \ref{Application} illustrates the application of our algorithm in both simulated and real data for SDSS AGNs in the Stripe 82 of the Sloan survey. Section \ref{LSST} discusses the potential impact of our methodology on the current LSST simulated data. Our conclusions are presented in the Section \ref{Summary}. We assume a flat $\Lambda$CDM cosmology with $H_0 = 69.8$ km s$^{-1}$ Mpc$^{-1}$, $\Omega_m = 0.28$. 

\section{Data}\label{data}
\subsection*{SDSS archival photometry data and light curves}

The SDSS provides observations in the $ugriz$ broad-band filters \citep{Gunn1998}, obtained using a dedicated 2.5m telescope \citep{Fukugita1996} at the Apache Point Observatory. Our initial sample consists of 9254 spectroscopically confirmed quasars with recalibrated light curves studied by M10 from the Stripe 82 (S82) region \citep{York2000,Annis2014}. Of these, 8974 quasars are drawn from the SDSS Data Release 5 (DR5) Quasar catalog \citep{Schneider2007}, with the rest taken from the Data Release DR7 \citep{Abazajian2009}
\footnote{\url{https://faculty.washington.edu/ivezic/macleod/qso_dr7/Southern.html}}.
S82 is the only stripe within the SDSS footprint that offers extensive multi-epoch imaging, making it suitable for AGN variability studies.
This region spans a $\SI{120}{\degree}$-long and $\SI{2.5}{\degree}$-wide stripe centered along the celestial equator. On average, this 290 $\text{deg}^2$ area has more than 60 epochs of observations per source for each filter. Since some observations were taken under non-photometric conditions, improved calibration methods have been applied to the sources in the S82 dataset by \citet{Ivezic2007} and \citet{Sesar2007}. These calibrations correct for systematic errors and improve the precision, which is particularly important for variability studies. These observations were done in annual ``seasons'', lasting approximately three months, for approximately 11 years, although, the earlier seasons have very sparse sampling ($\sim$ 1 to 3 observations per season). The final three observing seasons, covering a baseline of about 800 days, provide a denser sampling of more than 10 observations per season due to the SDSS supernova survey \citep{supernovasurvey}, which make them better suited for AGN variability studies. Therefore, our analysis is restricted to this well-sampled period taken later than Modified Julian Date (MJD) 53500 (after 2005). 

We used the 5-band ``BEST'' PSF photometry from the DR7 quasar catalog \citep{Schneider_2010}, calibrated following the procedure described by \citet{Richards2002}. These photometric data points are corrected for galactic extinction using the $u$-band ($A_u$) galactic extinction value from \citet{Schlegel1998}, while the galactic extinction values of $griz$ bands are $A_g, A_r, A_i, A_z = (0.736, 0.534, 0.405, 0.287)A_u$. To avoid the strong degeneracies in SED modeling at low redshifts, sources with $z<0.3$ are excluded, resulting in a refined parent sample of 9210 quasars. 
\begin{figure}[h]
    \centering
    \includegraphics[width=1\linewidth]{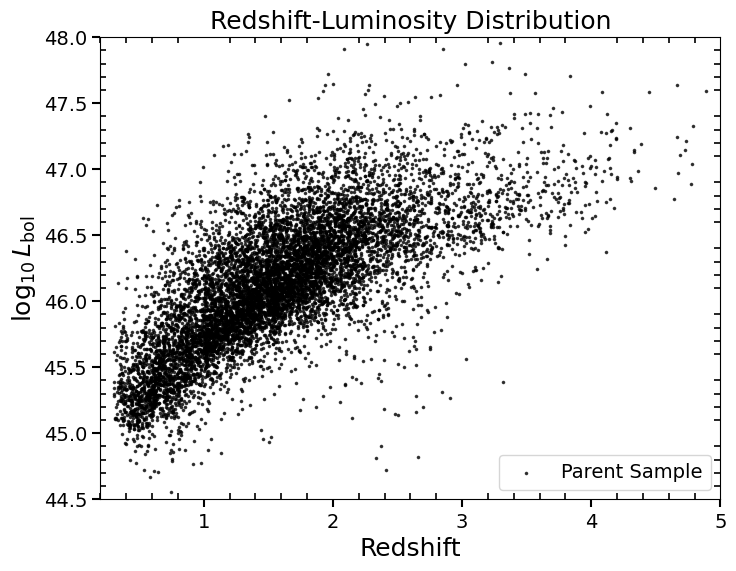}
    \caption{Distribution of the parent-sample quasars in the luminosity (bolometric) and redshift space. The bolometric luminosities ($ L_{\mathrm{bol}}$) of the quasars in this sample were estimated by \cite{Shen2011}.}
    \label{fig:Redshift_luminosity}
\end{figure}
Figure \ref{fig:Redshift_luminosity} shows the distribution of our parent sample in the redshift-luminosity plane. Note that only a small fraction of sources are found at $z<0.3$ (0.3\%), yet that is a part of the parameter space for SED modeling codes which can add significant degeneracies for the rest of the objects. For this reason, we limit our sample only to the 9210 quasars with $z>=0.3$.

\section{Methodology} \label{Methodology}

Previous studies \citep[e.g,][]{Collier_2001,Bauer2009}
confirm that the observed AGN variability can be statistically quantified using the Structure Function (SF). The SF is the Root Mean Squared (RMS) magnitude difference ($\Delta\text{m}$) as a function of the time difference ($\Delta\text{t}$) between observations, and is useful for analyzing sparsely or irregularly sampled light curves as it captures ensemble variability trends without relying on explicit modeling. For modeling AGN variability, the Damped Random Walk (DRW) process has been widely used to describe the SF. This process models the AGN variability as a stochastic process, defined by an exponential covariance matrix that behaves as a random walk for short timescales and asymptotically reaches a finite SF amplitude dubbed $\text{SF}_\infty$, at timescales which are comparable to or longer than the damping timescale ($\tau$) \citep{Kelly2009,Zu_etal2011,Ivezic2014,Suberlak_2021}. Although the physical origin is unclear, the DRW model can successfully describe many features of AGN light curves. For example, M10 modeled quasar light curves as a DRW process and studied whether the observed variability timescales of quasars could be linked to physical timescales in accretion disks and derived correlations between the best-fit parameters and the physical properties of the quasars. For a DRW process, the SF is described by
\begin{equation}
    \text{SF}(\Delta t)=  \text{SF}_\infty(1 - e^{-|\Delta t|/\tau})^{1/2}.
\end{equation}

The observed variability of quasars is governed by their intrinsic properties such as the black hole mass ($M_{\mathrm{BH}}$), the accretion rate, which is traced by the absolute $i$-band magnitude ($M_\mathrm{i}$), and the rest-frame wavelength ($\lambda_{\mathrm{RF}}$). Moreover, cosmological time dilation stretches the observed timescales and, consequently, the SF measured at a fixed observed-frame lag, making the variability appear redshift-dependent, although the asymptotic amplitude (SF$_\infty$) remains unaffected. Additionally, the measured SF$_\infty$ is modulated by the AGN flux fraction in each filter $j$, given by $R_{\mathrm{AGN}} = \frac{F^{j}_{\mathrm{AGN}}}{F^{j}_{\mathrm{AGN}} + F^{j}_{\mathrm{Galaxy}}}$, since the host galaxy contribution is non-variable.

\begin{figure}[h]
    \centering
    \includegraphics[width=\linewidth]{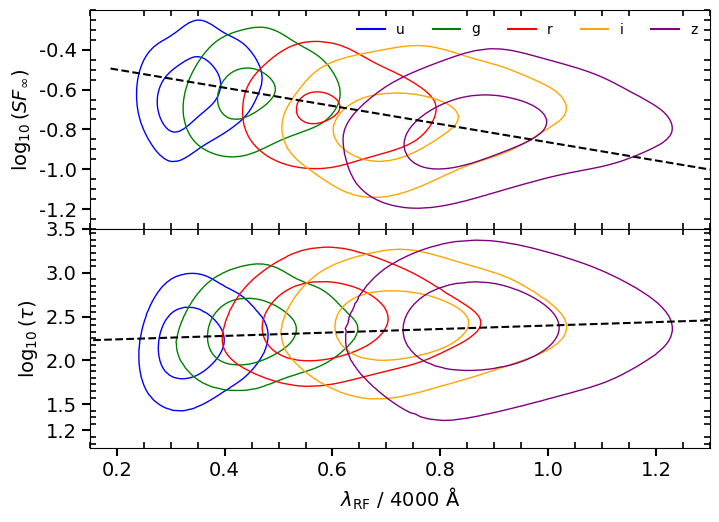}
    \caption{Distribution of the variability parameters, SF{$_\infty$} and $\tau$, in the observed frame, measured from the $ugriz$ bands as a function of rest frame wavelength. The coefficients of the Eq.~\eqref{equation} are calculated by fitting these values, corresponding to both the amplitude and timescale of variability. The dotted line represents the linear fit with slopes -0.456 and 0.19 for SF{$_\infty$} and $\tau$, respectively.}
    \label{fig:params_vs_lambda}
\end{figure}

To quantify how the DRW parameters vary with rest-frame wavelength ($\lambda_{\mathrm{RF}}$) and to derive the corresponding coefficients in our variability model outlined in this section,, we analyze their distributions across photometric bands. Figure~\ref{fig:params_vs_lambda} shows the distribution of the variability parameters SF{$_\infty$} and $\tau$ as a function of the $\lambda_{\mathrm{RF}}$, for each photometric band. While the scatter is large, the overall trend across filters is consistent with the expected decrease in variability amplitude at longer rest-frame wavelengths \citep{kubota2018} and the positive correlation in the characteristic timescale with longer rest-frame wavelengths \citep{2002Frank}. The model shown in Fig.~\ref{fig:params_vs_lambda} is included as a reference to illustrate the dependence of the parameters on $\lambda_{\mathrm{RF}}$. The amplitude is tailored to pass through the contours. The width of the contours is dependent on other physical parameters such as the $M_{\mathrm{i}}$, etc, that are not included in the model, and therefore appears broader than the scatter associated with the wavelength dependence alone. As a result, the contour widths should not be interpreted as uncertainties of the fit. Note that while the correlation in $\tau$ is significant, it is hard to observe in the Figure owing to the wide dynamic range spanned by the measurements. The results from M10 showed that, even though $\tau$ and SF{$_\infty$} exhibit weak trends with redshift and luminosity, the most significant correlation are with $M_{\mathrm{BH}}$ and Eddington ratio ($L_\mathrm{bol}/L_{\rm Edd}$) (see their Figure 15), which is consistent with the thermal or viscous disk timescales. They accounted for the uncertainties in $M_{\mathrm{BH}}$ ($\sim0.4$ dex; \citealp[]{yueshen2024}) and demonstrated that ignoring this uncertainty can lead to underestimation of the $M_{\mathrm{BH}}$ dependence. Additionally, since the $L_\mathrm{bol}/L_{\rm Edd}$ depends on both $L_\mathrm{bol}$ and $M_{\mathrm{BH}}$, it is probable that the observed trends in SF$_\infty$ reflect an underlying dependence on the $L_\mathrm{bol}/L_{\rm Edd}$ rather than luminosity or mass alone. However, because $M_{\mathrm{BH}}$ estimates are often unavailable or uncertain in large surveys, they also provided reduced models excluding it. Furthermore, DRW is a stationary Gaussian process, and \citet{Stone2022} showed that when the baseline extends to $\sim$20 years, the best-fit $\tau$ continues to increase, suggesting that AGN variability may not be strictly stationary over multi-year or decade timescales. Thus, the DRW model should be regarded as a convenient simplification. 
The DRW parameters $\tau$ and SF{$_\infty$} are modeled by M10 as
\begin{align}
    \log f = A + B \log \left( \frac{\lambda_{\text{RF}}}{4000\ \mathrm{\AA}} \right) + C (M_\mathrm{i} + 23) \notag \\
           \quad + D\log \left(\frac{M_{\mathrm{BH}}}{10^9 M_\odot}\right) + E \log(1+z),
    \label{equation_original}
\end{align}
\noindent where $f$ is either SF{$_\infty$} or $\tau$, $M_{\mathrm{i}}$, and $A, B, C , D$, and $E$ are fit to the ensemble of objects. In our specific case, to model the AGN variability, we used the best-fit relation which does not have an explicit dependence over the $M_{\mathrm{BH}}$, resulting in a simplified form solely dependent upon $M_\mathrm{i}$ and $\lambda_{\mathrm{RF}}$. Specifically, we follow M10 and set $D=0$ and $E=0$, such that
\begin{equation}
    \log_{10} f = A + B \log_{10}\!\left( \frac{\lambda_{\mathrm{RF}}}{\SI{4000}{\angstrom}} \right) 
    + C \,(M_{\mathrm{i}} + 23).
    \label{equation}
\end{equation}

Note that although we used Eq.~\eqref{equation} for the modeling to estimate redshift priors through \texttt{VAR-PZ} (see below), the relation (Eq.~\ref{equation_original}) which incorporates the dependency on the $M_{\mathrm{BH}}$ was used only to simulate the realistic SDSS and LSST light curves (see Section \ref{Simulation}). The $A, B$ and $C$ coefficients in this work are re-estimated to overcome the potential biases due to different algorithms for light curve modeling and in estimating the AGN continuum luminosities. Specifically, the $M_\mathrm{i}$ and $R_{\mathrm{AGN}}$ values are obtained through SED modeling of the SDSS photometry using the Low Resolution Templates (\texttt{LRT}) and algorithm of \citet[][hereafter A10, see Section \ref{Code}]{Assef2010}. We find the best-fit values of the $A, B$ and $C$ coefficients in Eq.~\eqref{equation} following a similar approach to M10. We first estimate $B$ by fitting a power-law of the form 
$f \propto \left( \frac{\lambda_{\mathrm{RF}}}{\SI{4000}{\angstrom}} \right)^{B}$
for both SF{$_\infty$} and $\tau$ parameters to every quasar with at least two observations in multiple bands. We adopt as the value of $B$ for all subsequent calculations as the median of all objects for which we were able to carry out this fit, corresponding to $-0.456 \pm 0.03$ and $0.19 \pm 0.01$ for SF{$_\infty$} and $\tau$, respectively (see Sect. 5.1 in M10). We then fix $B$ and fit the values of coefficients $A$ and $C$ to the full ensemble of objects. This method of fixing coefficient $B$ before other coefficients is done to eliminate the degeneracies between the $\lambda_{\text{RF}}$ and other physical parameters, as $B$ is the only parameter in a term that depends on wavelength. This process effectively decouples $B$ from the rest of the parameters and minimizes the covariances.

\begin{table*}[t]
\centering
\renewcommand{\arraystretch}{1.2}
\caption{Best-fit coefficients for Eq.~\eqref{equation} from M10 and this work.}
\begin{tabular}{l llll l}
\toprule
\textit{f} & \textit{A} & \textit{B} ($\lambda_{\text{RF}}$) & \textit{C} ($M_{\text{i}}$) & \textit{D} ($M_{\text{BH}}$) & Reference \\
\midrule \midrule
SF$_{\infty}$ & $-0.51 \pm 0.02$ & $-0.479 \pm 0.005$ & $0.131 \pm 0.008$ & $0.18 \pm 0.03$ & M10 \\
SF$_{\infty}$ & $-0.618 \pm 0.007$ & $-0.479 \pm 0.005$ & $0.090 \pm 0.003$ & 0   \\
$\tau$        & $2.4 \pm 0.2$ & $0.17 \pm 0.02$ & $0.03 \pm 0.04$ & $0.21 \pm 0.07$   \\
$\tau$        & $2.2 \pm 0.1$ & $0.17 \pm 0.02$ & $-0.01 \pm 0.02$ & 0   \\
\midrule
SF$_{\infty}$ & $-0.695 \pm 0.001$ & $-0.456 \pm 0.03$ & $0.086 \pm 0.001$ & $0.012 \pm 0.001$ & This Work \\

SF$_{\infty}$ & $-0.713 \pm 0.003$ & $-0.456 \pm 0.03$ & $0.071 \pm 0.002$ & 0   \\
$\tau$        & $2.2 \pm 0.004$ & $0.19 \pm 0.01$ & $-0.027 \pm 0.002$ & $0.027 \pm 0.003$ &  \\

$\tau$        & $2.2 \pm 0.019$ & $0.19 \pm 0.01$ & $-0.026 \pm 0.011$ & 0   \\

\bottomrule
\end{tabular}
\tablefoot{Coefficients are provided for both variability parameters, SF$_{\infty}$ and $\tau$, with and without $M_{\mathrm{BH}}$ dependency. The recalibration of these coefficients is done to avoid inconsistencies arising from using different $M_{\mathrm{i}}$ estimates and light curve modeling algorithms.}
\label{tab:variability_params}
\end{table*}

Table \ref{tab:variability_params} shows the values of the coefficients $A$, $B$, $C$ and $D$ of the SF{$_\infty$} and $\tau$ relations in  Eq.~\eqref{equation_original}, respectively, in comparison to the values estimated by M10. The small variations in the coefficients with respect to those from M10 are likely due to methodological changes in both light curve modeling and $M_\mathrm{i}$ estimation.
\begin{figure*}    
    \centering
    \includegraphics[width=\textwidth, trim=0 8 0 6, clip]{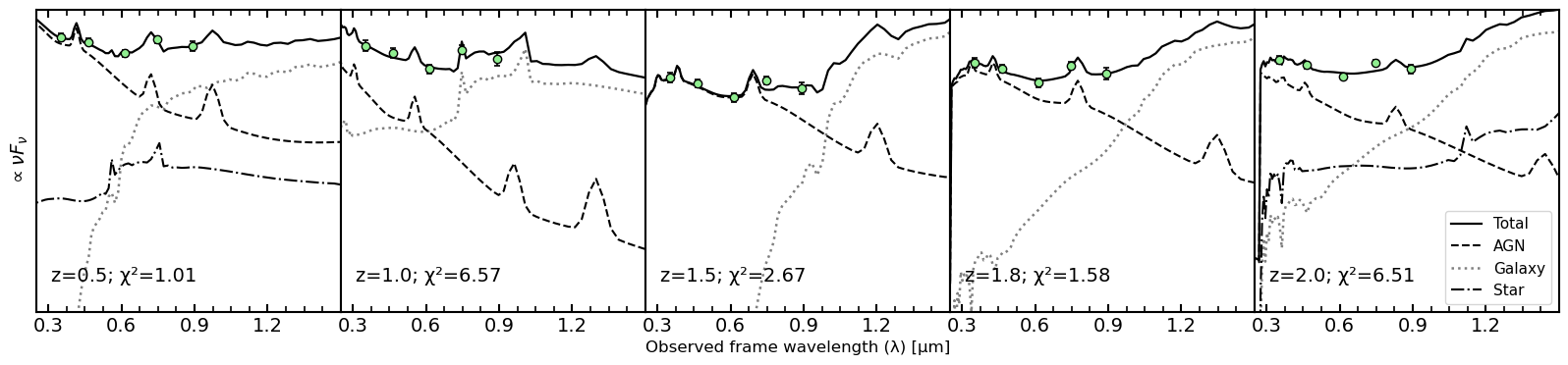}
    
    \vspace{0.5em}
    \includegraphics[width=\textwidth, trim=0 8 0 6, clip]{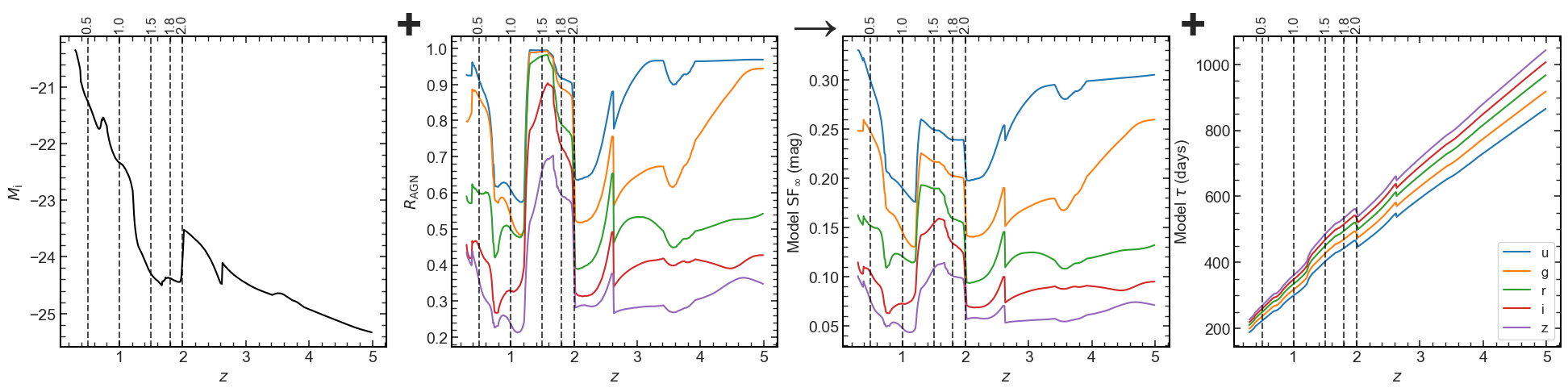}
    
    \vspace{0.5em}
    
    \hspace*{-0.01\textwidth}
    \includegraphics[width=1.02\textwidth, trim=0 10 80 6, clip]{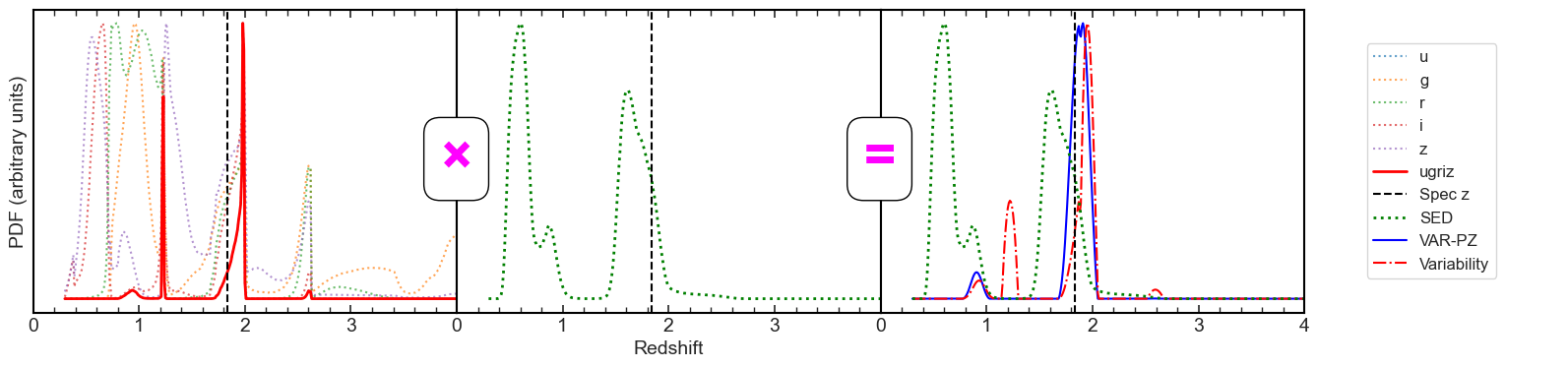}
    
    \label{fig:PDF_explanation}
    
    \caption{ \textit{Top panel:} SED fits at trial redshifts z = 0.5, 1.0, 1.5, 1.8, and 2.0 with corresponding $\chi^2$ values. \textit{Middle panel:} \texttt{VAR-PZ} redshift prior estimation. Left: 
    $M_{\mathrm{i}}$ and $R_{\mathrm{AGN}}$ components estimated from the SED modeling as a function of redshifts. Right: Expected 
    SF$_{\infty}$ and $\tau$ using the Eq.~\eqref{equation} for the corresponding redshifts. \textit{Bottom panel:} Photo-$z$ priors from \texttt{VAR-PZ} along with SED fitting PDF, and the combined posterior. Vertical dashed lines mark the trial redshifts from the top panel.}
\label{fig:PDF_explanation}
\end{figure*}

In order to model the AGN variability as a DRW process and to obtain the photo-$z$ prior, we used Gaussian Process (GP) along with the Maximum Likelihood Estimation (MLE) framework. GP is a flexible non-parametric method that can model the covariance structure of time series data and MLE is a statistical approach for estimating the likelihood of the fitted parameters, respectively. We use Eq. ~\eqref{equation} to calculate the pair of DRW parameters (SF{$_\infty$} and $\tau$) as a function of redshift for each photometric band using the $M_{\mathrm{i}}$ and $R_{\mathrm{AGN}}$ values obtained from the SED modeling at those redshifts. These redshift-dependent DRW parameters are used as inputs to the GP-based DRW model, where for each trial redshift, the corresponding DRW parameters are applied to fit the light curve in each photometric band. The MLE quantifies how likely are those DRW parameters to result in the observed light curve, providing the likelihood of that trial redshift. The resulting likelihood distributions from all bands are then multiplied to construct a prior PDF for constraining photo-$z$s. This PDF (hereafter \texttt{VAR-PZ} PDF) is then combined with the PDF derived from the SED modeling (hereafter \texttt{SED-PDF}) to produce a more accurate posterior probability distribution. This approach enables us to evaluate the contribution of the \texttt{VAR-PZ} PDF in improving the photo-$z$ predictions.

Figure~\ref{fig:PDF_explanation} schematically illustrates a characteristic example of this approach for an S82 quasar (``SDSS J000013.80-005446.8'', $z_{\text{spec}} = 1.8361$), demonstrating how \texttt{VAR-PZ} constrains the photo-$z$ estimation. The top panel shows the SED fitting results to the stacked SDSS photometry for different trial redshifts (0.5, 1.0, 1.5, 1.8, and 2.0), with each redshift represented by a different color along with their corresponding $\chi^2$ fit values. Note that for different redshifts, the best-fit AGN luminosities (parametrized by $M_\mathrm{i}$) and the fraction of the flux coming from the AGN component ($R_{\mathrm{AGN}}$) can vary significantly, which in turn affects the expected variability properties (see Section \ref{Code}). In this example, the SED fitting strongly favors a low-redshift solution. The middle panel presents the steps occurring within \texttt{VAR-PZ} for constraining the photo-$z$s. The left panels show the $M_{\mathrm{i}}$ and $R_{\mathrm{AGN}}$ values estimated through the SED modeling as a function of redshift, while the right panels show the expected SF$_{\infty}$ and $\tau$ values expected as a function of redshift to estimate the DRW model likelihood. When this technique is applied across the full redshift grid, it generates the PDFs shown in the bottom panel. Vertical dashed lines mark the trial redshifts from the top panel. The first subplot shows \texttt{VAR-PZ} PDF across all five bands, which strongly favors higher redshifts. The second subplot displays the PDF derived solely from SED fitting, which peaks at low redshift. The third subplot presents the posterior obtained through the combination of the SED and variability PDFs. This combined approach effectively discards the low-redshift solutions preferred by SED fitting alone, with the variability constraints driving the posterior toward the true spectroscopic redshift. This example demonstrates how \texttt{VAR-PZ} breaks degeneracies inherent in SED-only photo-$z$ estimation and improves redshift accuracy.

\section{Implementation}\label{Code}

In this section, we explain the \texttt{VAR-PZ} routine, detailing how the code models AGN variability using the DRW process and incorporates it into the photo-$z$ estimation.
We employed the Python-based Gaussian Process (GP) toolkit \texttt{Celerite} \citep{Foreman-Mackey2017} to model AGN light curves as a DRW process. \texttt{Celerite}'s computationally efficient GP implementation enables faster modeling compared to traditional methods such as \texttt{Javelin} \citep{javelin2010}. 
It provides the flexibility to use MLE or Markov Chain Monte Carlo (MCMC), depending on our specific needs. Although, MCMC provides a thorough exploration of parameter space, it is often computationally more expensive compared to methods which can leverage MLE.
We developed a custom DRW kernel to use \texttt{Celerite} for our goals. Unlike the default \texttt{Celerite} kernel (\texttt{RealTerm}) used for DRW fitting, which models the SF{$_\infty$} and $\tau$ independently, our kernel jointly fits for both of them using the relations outlined in Eq.~\eqref{equation} with the coefficients in Table \ref{tab:variability_params}. \footnote{While \texttt{Celerite} internally models variability using the variance parameter $\sigma^2$, we apply a transformation to recover SF{$_\infty$} via the relation SF{$_\infty$} = $\sqrt{2} \sigma$, as commonly adopted in variability analyses.}

The following stage of the implementation details how the kernel integrates the redshift-dependent DRW parameters into the \texttt{VAR-PZ} framework to generate the variability prior. The kernel involves two main stages. An initialization computes the value of $M_\mathrm{i}$ for the AGN component as well as $R_{\mathrm{AGN}}$ to account for the host fractions in the photometric band of the light curve being modeled, as a function of redshift in a grid between $z$=0 and $z$=5 with a bin size of 0.01. These values are obtained by modeling the observed SED adopting each redshift of the grid. We nominally interpolate between redshifts for flexibility, but in practice we minimize the use of interpolation by matching the redshift grids of the \texttt{VAR-PZ} and SED PDFs. In the next stage, the kernel computes the expected SF{$_\infty$} and $\tau$ values based on the estimated $M_\mathrm{i}$ values for each band in the light curve, using Eq.~\eqref{equation}. The value of $\tau$ is then modified to account for cosmological time dilation while the SF{$_\infty$} is scaled by the AGN flux fraction ($R_{\mathrm{AGN}}$) to account for the contribution of the host galaxy. This renormalization approach helps to account for the damped variability observed in intermediate-type AGN where host galaxy contamination significantly affects the measured variability amplitude. While this work does not explicitly model the physical parameters of the host galaxy and their connection the AGN properties, future refinements could incorporate additional host galaxy properties, such as the $M_{\mathrm{BH}} - M_{*}$ relation, which influences the $L_\mathrm{bol}/L_{\mathrm{Edd}}$, which could potentially provide stronger constraints on the \texttt{VAR-PZ} PDF.

We then compute the likelihood distribution as described earlier, independently for each band and combine them to produce the \texttt{VAR-PZ} priors. To account for additional uncorrelated white-noise components in the light curves, we included a \texttt{JitterTerm} kernel in our model. This term reduces the effect of photometric noise on the AGN variability parameters fit. We describe the workflow in detail in Appendix \ref{fig:routine_flowchart}. We have made our implementation of the method publicly available here\footnote{%
\url{https://github.com/SarathSS98/VAR-PZ/tree/main} \\
}
 along with the required documentation.

\section{Application to SDSS quasars}\label{Application}
As a proof of concept, we apply our methodology to SDSS quasars in S82, first with simulated light curves and then with the observed light curves to estimate the improvement in photo-$z$s obtained by using the \texttt{VAR-PZ} priors. To assess performance of our photo-$z$s, the outlier fraction and precision of the estimates are calculated. Following \citet{salvato2019many}, we define outliers as objects for which
\[
\frac{|z_{\text{phot}} - z_{\text{spec}}|}{1 + z_{\text{spec}}} > 0.15
\]

\noindent where $z_{\text{phot}}$ is the photometric redshift estimate, and $z_{\text{spec}}$ is the spectroscopic redshift. 
We evaluate the precision of the estimates through the Normalized Median Absolute Deviation (NMAD; e.g, \citealp{salvato2019many}) of the non-outlier objects. The NMAD is defined as
\[
\sigma_{z_p} = 1.48 \times \text{median} \left( \frac{|z_{\text{phot}} - z_{\text{spec}}|}{1 + z_{\text{spec}}} \right).
\]

\subsection{Simulated light curves}\label{Simulation}
We start by looking at the simplest scenario by simulating the light curves of our parent sample of S82 quasars assuming a DRW model and Eq.~\eqref{equation_original}. Note that, we use this Eq.~\eqref{equation_original} that incorporates the $M_\mathrm{BH}$ dependency to simulate the light curves rather than Eq.~\eqref{equation}, which is the one used to create the \texttt{VAR-PZ} priors, as it should provide a better approximation to the real SDSS light curves (see M10 for details). The light curves are simulated through \texttt{Celerite} using the implementation of \citet{Burke2021} for a grid of observing seasons and cadences, incorporating realistic photometric uncertainties and realistic seasonal observing gaps. The simulated light curves span a wide grid of cadence intervals ranging from 1 to 10 days, and baseline lengths ranging from 1 to 10 seasons. Each S82 observing season was assumed to last 90 days, with seasonal gaps of 275 days. Realistic SDSS photometric uncertainties are added to our simulated light curves according to the brightness of each target. This approach enables an isolation of the effect of varying cadences and baselines.
\begin{figure}[h]
    \centering
    \includegraphics[width=1\linewidth]{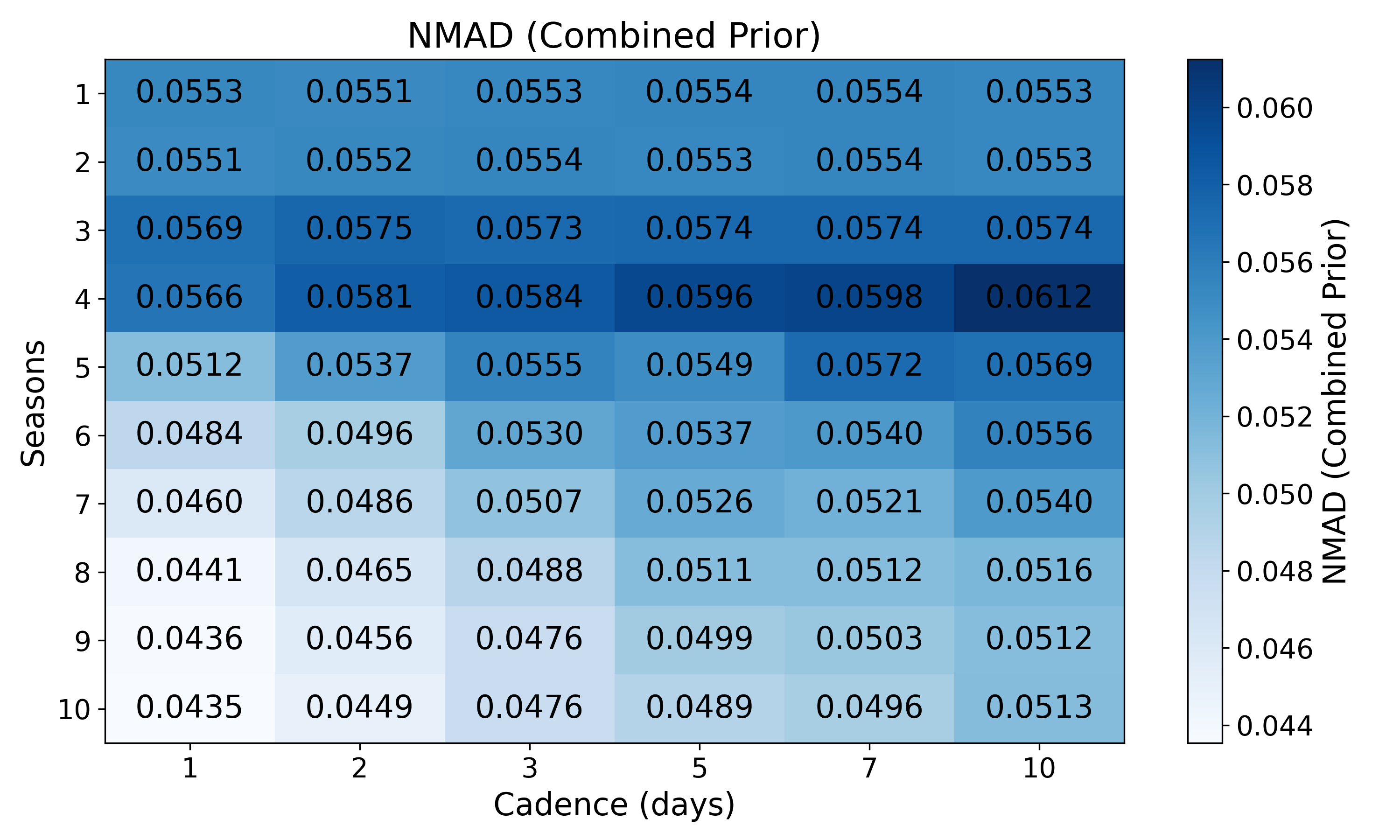}
    \includegraphics[width=1\linewidth]{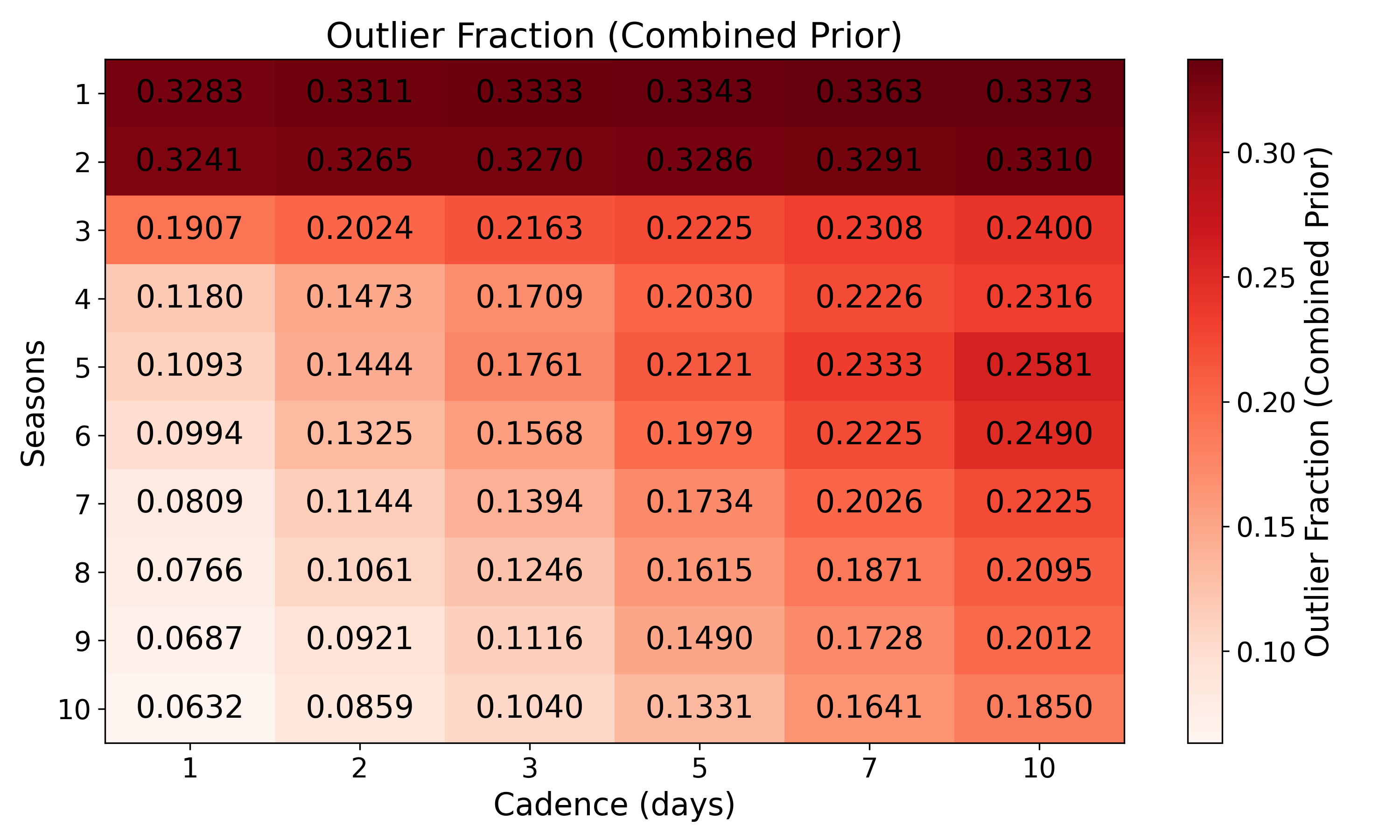}
    \caption{Heatmap illustrating the influence of observational cadence and baseline length (in SDSS seasons) on photometric redshift estimation metrics: the NMAD (top) as a measure of precision, and the outlier fraction (bottom). The color bar, presented on a logarithmic scale, maps these metrics such that regions depicting lower values correspond to the most promising observing conditions, yielding higher precision and a reduced fraction of outliers in photometric redshift predictions.}
    \label{heatmap}
\end{figure}

According to \citet{Suberlak_2021}, accurate recovery of the damping timescale, $\tau$ requires a light curve baseline at least three times longer than this timescale. Therefore, depending on the length of the baseline, \texttt{VAR-PZ} priors are constructed using the method outlined earlier in the range of redshifts where the baselines are at least three times longer than the expected value of $\tau$. The remaining part of the PDF is simply flat, with a value given by the maximum of the PDF, where the $\mathrm{baseline} \geq 3\tau$. This strategy enables an efficient application of the variability constraints only in the part of the parameter space where there is constraining power for our method.
 
Figure \ref{heatmap} presents a heatmap, which quantifies the impact of priors generated in this range of the baselines and observational cadences on both the outlier fraction and redshift precision of the photometric redshifts obtained using \texttt{LRT} (see Section~\ref{Methodology}). For reference, without using priors, \texttt{LRT} obtains $\sigma_{zp}$ = 0.0554 and an outlier fraction of 0.33. As expected, the largest improvements are achieved with a combination of long baselines and high observational cadences. While the improvements in the precision are modest, the reduction in the outlier fraction is substantial. In fact, for the highest cadence and largest number of seasons probed, \texttt{VAR-PZ} is able to constrain photo-$z$s, with an accuracy ($\sigma_{zp}$) = 0.0435 and an outlier fraction of 0.06. 

In contrast, with limited temporal coverage, specifically, fewer than two seasons and a cadence $\geq 7$ days, the addition of \texttt{VAR-PZ} does not improve upon SED-only methods for redshift estimation, and thus requires careful application to avoid degrading performance.
The diagonal structure of the heatmap reveals a partial degeneracy between cadence and baseline; that is, higher cadence can partially compensate for shorter baselines and vice versa, underlining potential observational strategies. 

\subsection{Observed light curves}\label{SDSS_real}
We applied our methodology to the real SDSS Stripe~82 light curves (see Section \ref{data}). To ensure the use of reliable and high-quality data,
the light curves were filtered to exclude flux measurements with a signal-to-noise ratio ($\mathrm{S/N}$) below 3 relative to the zero–flux level, thereby ensuring that the median flux of each light curve is not biased by measurements at the faint end.\footnote{\url{https://www.sdss4.org/dr17/algorithms/magnitudes/}}. Using this median as a reference, we then exclude observations which
deviate by more than $2.5\sigma$, thereby eliminating strong outliers. Finally, a quality cut is applied by retaining only epochs with $\mathrm{SNR} > 10$. This filtering ensures that only high-quality measurements are included in the light curves, though we note that they may also introduce biases in the measured light curve properties. We only use light curves for a given band to contribute to the \texttt{VAR-PZ} prior if they have more than 35 measurements (which corresponds to the peak of the data point distribution in the light curves). Out of 9210 total sources, 6493 met these criteria in all five bands, and, 8413 do so in at least one band. We remove the remaining 797 sources from our analysis. As discussed in the previous section, the relatively short temporal baselines of the SDSS light curves restrict their constraining power at higher redshifts. 
\begin{figure}[h]
    \centering
    \includegraphics[width=1\linewidth]{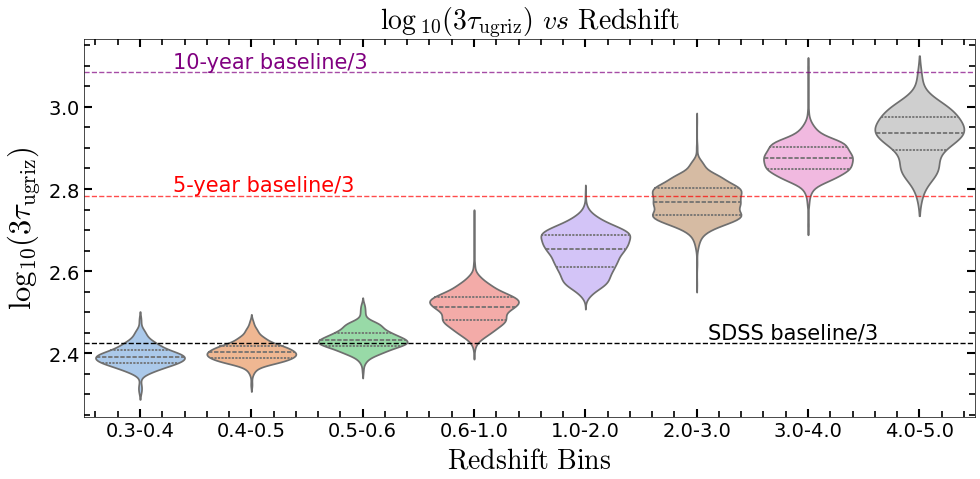}
    \caption{Violin plot showing the median AGN variability timescale ($\tau$) estimated from M10 over the SDSS bands as a function of redshift. The dashed black line represents one-third of the SDSS baseline duration, showing the redshift-dependent threshold beyond which the condition $3\tau<\mathrm{baseline}$ is no longer satisfied. This requirement defines the redshift range over which \texttt{VAR-PZ} can constrain redshifts with the current data. The red and purple line represent one-third of the 5 and 10 year LSST baseline, respectively. LSST's enhanced sensitivity will enable the detection of fainter objects, resulting in lower overall timescale values ($\tau$) and consequently providing greater constraining power.}
    \label{zs_vs_tau}
\end{figure}
Figure~\ref{zs_vs_tau} illustrates the distribution of median $\tau$ values, estimated using Eq.~\eqref{equation_original}, as a function of redshift for our AGN sample. The black dashed horizontal line represents one-third of the SDSS S82 baseline length. This highlights a redshift-dependent limitation: beyond $z\sim0.5-0.6$, the S82 data do not provide a sufficiently long baseline to reliably constrain the DRW parameters, and consequently, the accuracy of the redshift (VAR-PDF) priors diminishes at higher redshifts. Hence, we followed the same implementation strategy mentioned before to produce the posterior distribution function, namely, we replace all likelihood values for redshifts where the expected $\tau$ from Eq.~\eqref{equation} exceeds on third of the baseline for the maximum likelihood observed for those points. The red and purple lines in Figure \ref{zs_vs_tau} represent one-third of the 5-year and 10-year LSST baseline, respectively, compared to these objects, although note that the higher photometric sensitivity of LSST will extend AGN detection to much fainter flux limits compared to SDSS, systematically shifting the population toward lower $\tau$ values. The increased survey depth will yield improved constraints for more AGNs even when considering similar shorter observational baselines in the initial years of the survey. 

 \begin{figure}[h]
    \centering
    \includegraphics[width=1\linewidth]{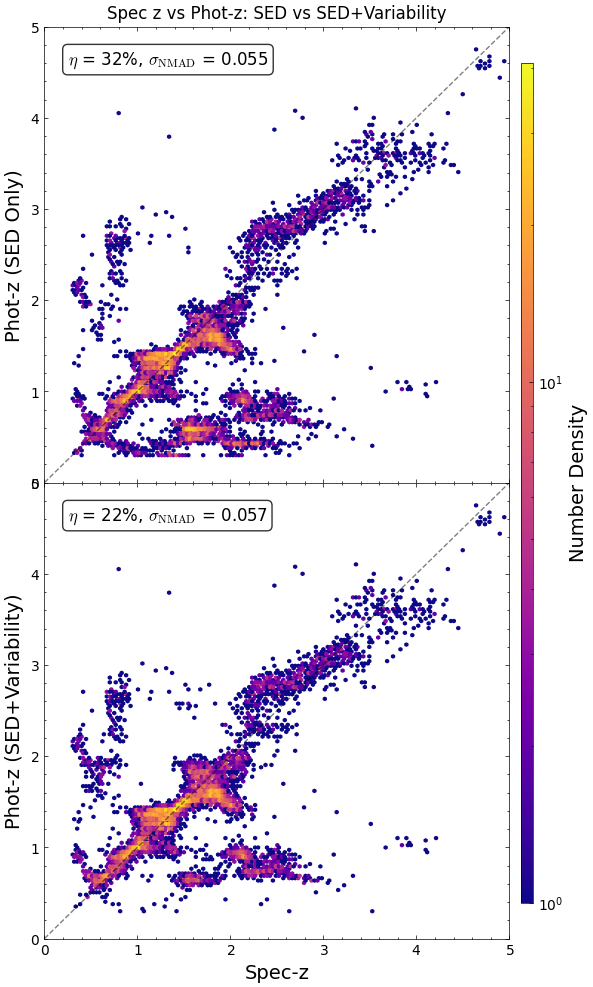}
    \caption{Binned scatter diagrams comparing photometric redshifts derived from SED fitting using \texttt{LRT}, independently (top) and combined with our variability model (bottom), against spectroscopic redshifts. The dashed line represents the one-to-one correspondence between the axes.}
    \label{fig:zp_vs_zs_real}
\end{figure}

 \begin{figure}[h]
    \centering
    \includegraphics[width=1\linewidth]{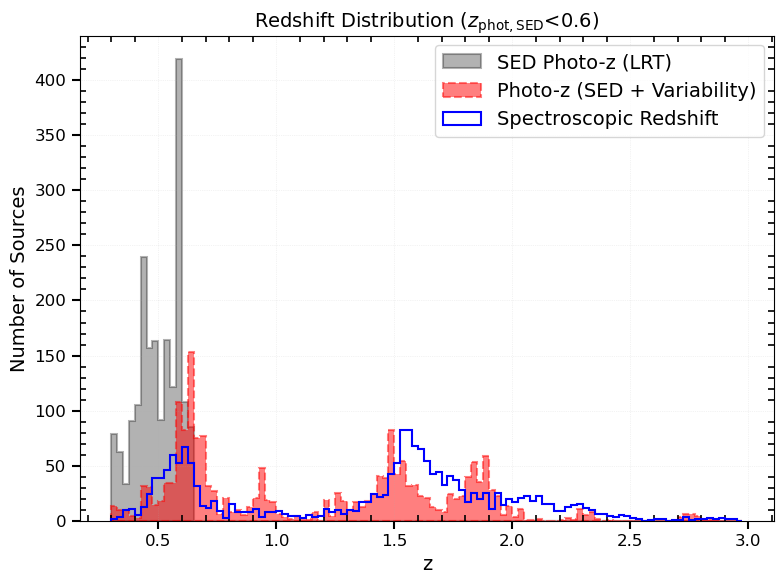}
    \caption{Distribution of photometric redshifts for sources with $z_{\mathrm{phot}}<0.6$, comparing \texttt{LRT} SED-only (gray) and \texttt{VAR-PZ}-enhanced (red) solutions, illustrating how variability priors redistribute sources and remove catastrophic photo-$z$ failures in this regime. The blue histogram represents the spectroscopic redshifts of those sources.
    }
    \label{fig:low_pz}
\end{figure}

Figure \ref{fig:zp_vs_zs_real} compares the photometric and spectroscopic redshifts before and after applying the variability priors. The inclusion of \texttt{VAR-PZ} priors significantly improves the SED-derived probability distributions by reducing catastrophic failures and resolving degeneracies inherent to SED fitting alone. The accuracy estimates of this method on the SDSS observations are shown in Table \ref{tab:photz_performance}. For our sample, the inclusion of variability reduces the outlier fraction from 32\% to 22\%. This effect is particularly pronounced in the regime where SED fitting predicts low redshifts ($z_{\mathrm{SED}} < 0.6$), where variability plays a critical role in mitigating catastrophic failures: the outlier fraction in this regime drops from 81\% to 33\%, as shown in Figure \ref{fig:low_pz}. Note that, as described in Section~\ref{data}, we restricted our analysis to sources with $z > 0.3$ to avoid significant degeneracies arising from the \texttt{LRT} SED modeling. Although this removes the low-redshift regime, it represents only $\sim$0.3\% of the parent sample and therefore has a negligible impact on the overall statistics. Retaining these low-$z$ sources would lead to the overestimation of the \texttt{VAR-PZ}'s performance, as the \texttt{LRT} SED modeling algorithm tends to produce degenerate redshift solutions in this low-$z$ regime, resulting in an outlier fraction of $\sim$47\% for the SED-only case. Excluding them thus ensures that the reported improvement of \texttt{VAR-PZ} accurately reflects its true performance rather than being driven by SED-induced degeneracies.

As an independent test, we study the impact of the \texttt{VAR-PZ} approach by applying its priors to the SED PDFs computed by the more sophisticated \texttt{LePHARE} SED fitting framework \citep{Arnouts1999, Ilbert2006} which has been optimized for AGNs by \citep{Shirley_submitted} to estimate the improvements. Note, however, that the computation of $M_\mathrm{i}$ and $R_{\mathrm{AGN}}$ for \texttt{VAR-PZ} is still computed using the \texttt{LRT} SED modeling as the host-AGN decomposition is not currently computed by \texttt{LePHARE}. When applying the VAR-PZ priors, the outlier fraction decreases from 23\% to 19\% overall for the sample, while for sources where \texttt{LePHARE} predicts a $z<0.6$ the outlier fraction improves from 66\% to 37\%, further demonstrating the robustness of variability as a complementary prior. We note that a, more pronounced improvement by the addition of \texttt{VAR-PZ}, potentially even comparable to those seen in \texttt{LRT}, could be possible, if the $M_\mathrm{i}$ and $R_{\mathrm{AGN}}$ values were derived self-consistently with \texttt{LePHARE}.
While introducing luminosity function (LF) priors on the host galaxies (see A10 for details on the implementation) can overcome some of the degeneracies resolved by \texttt{VAR-PZ}, they often compromise the accuracy of redshift estimates for sources that are otherwise well-constrained, highlighting the need for a more sophisticated LF prior which closely matches the distribution of quasar host galaxies. 

The short baseline of the SDSS observations limits the achievable improvements, so in the next section we study the potential impact when considering the observations of the upcoming Rubin LSST, which will provide longer and denser light curves spanning a decade \citep{Ivezic2019}. Note that while, in principle, data from Zwicky Transient Facility (ZTF) could be used to extend the temporal baseline of the sources in S82, their relatively large photometric uncertainties (with a median value of about $\sim0.1$ mag for the objects in our sample) compared to SDSS ($\sim0.02$ mag) result in only small improvements. In a similar context to our study, \citet{Suberlak_2021} found that increased noise effectively offsets the potential benefits of the longer time coverage.

\begin{table*}[h]
\caption{Comparison of photo-$z$ performance for SED and SED + variability models.}
\centering
\renewcommand{\arraystretch}{1.2}
\setlength{\tabcolsep}{12pt} 
\begin{tabular}{l c c c c c} 
\toprule
\multicolumn{6}{c}{\textbf{Total Sources}} \\  
\midrule
\textit{Method} & 
\textit{\shortstack{Number \\ of Sources}} & 
\textit{\shortstack{Number \\ of Outliers}} & 
\textit{\shortstack{Outlier \\ Fraction}} & 
\textit{\shortstack{NMAD Without \\ Outliers}} & 
\textit{\shortstack{NMAD With \\ Outliers}} \\  
\midrule
LRT SED & 8413 & 2745 & 0.3263 & 0.0558 & 0.1051 \\  
LRT SED + Variability & 8413 & 1918 & 0.2280 & 0.0579 & 0.0850 \\  
LePHARE SED & 8413 & 1954 & 0.2323 & 0.0483 & 0.0705 \\  
LePHARE SED + Variability & 8413 & 1642 & 0.1952 & 0.0485 & 0.0663 \\  
\midrule
\multicolumn{6}{c}{\textbf{$z_{\mathrm{SED}}<0.6$}} \\  
\midrule
LRT SED & 1300 & 1060 & 0.8154 & 0.0410 & 0.6105 \\  
LRT SED + Variability & 1300 & 429 & 0.3300 & 0.0579 & 0.1077 \\  
LePHARE SED & 734 & 491 & 0.6689 & 0.0379 & 0.6059 \\  
LePHARE SED + Variability & 734 & 278 & 0.3787 & 0.0499 & 0.1242 \\  
\bottomrule
\end{tabular}
\tablefoot{ NMAD represents the Normalized Median Absolute Deviation. Values are shown for the entire AGN sample and for the subsample with SED photo-$z$s less than 0.6.}
\label{tab:photz_performance}
\end{table*}

\section{Prospects for Rubin LSST}\label{LSST}
Rubin LSST is expected to revolutionize our understanding of AGNs. With between 50 and 200 visits per source in each of the ``ugrizy'' filters, the ten-year survey will observe at least 10 million AGNs spanning 18,000 $\mathrm{deg}^2$ of the sky up to redshift $ z\sim7.5$, with average
luminosities of  $L_{\rm bol} \sim 10^{44}\ \mathrm{erg\,s^{-1}}$ (\citealt{lsstsciencebook,Li_2025}).\footnote{\url{http://lsst.org}}. We also expect to detect an additional $\sim 40,000$ fainter AGNs with more than 1000 samplings per band in the Deep Drilling Fields (DDFs), a smaller sky area ($60$ deg$^2$) with very high cadence \citep{brandt2018activegalaxysciencelsst}. LSST's long-term monitoring will provide a complete census of AGNs, and its exceptional cadence with small photometric uncertainties, will greatly enhance variability studies. This coverage should be ideal for \texttt{VAR-PZ} to improve the accuracy of the redshift predictions.

 We used the expected bandpasses provided by the observatory\footnote{{\url{https://github.com/lsst/throughputs/tree/main/baseline}}} to simulate realistic photometry.
The synthetic stacked LSST photometry were simulated by using the best-fit SED model to the observed SDSS $ugriz$ photometric data obtained with \texttt{LRT} from A10 at their corresponding spectroscopic redshifts convolved with the LSST bandpasses. \cite{Ivezic2019} provide comprehensive LSST performance measurements. The anticipated photometric error in magnitudes for a single visit can be expressed as
\begin{equation}
 \sigma_{\text{source}} = \sqrt{\sigma_{\text{rand}}^2 + \sigma_{\text{sys}}^2} 
\end{equation}
where $\sigma_{sys}$ is the systematic photometric error (caused, for instance, by imprecise PSF modeling, but excluding uncertainties in the absolute photometric zero-point) and $\sigma_{\text{rand}}$ is the random photometric error (see \citet{Ivezic2019} for details). The survey's calibration system is developed to limit the systematic error to $\sigma_{\text{sys}} < 0.005$ mag. For point sources, \citet{Ivezic2019} adopts the following expression based on \citet{Sesar2007} for the $\sigma_{\text{rand}}$ in magnitudes,
\begin{equation}
    \sigma_{\text{rand}}^2 = (0.04 - \gamma)x + \gamma x^2
\end{equation}
where $\mathrm{log_{10}}(x)= 0.4(m - m_{5s})$ with $m$ being the observed magnitude and $m_{5s}$ being the $5\sigma$ depth of the coadded magnitude limit of the LSST filters\footnote {\url{https://rubinobservatory.org/for-scientists/rubin-101/key-numbers}}.
 The parameter $\gamma$ is filter-dependent, with $\gamma_u = 0.038$ for the $u$-band and $\gamma = 0.039$ for the $g$, $r$, $i$, $z$, and $y$ bands \citep{Ivezic2019}. Note that, LSST observations in different bands are not simultaneous, and the associated variability-induced color offsets are not included in our simulations. Realistic LSST light curves are simulated for the parent sample using the current LSST baseline v5.0.0 of the Operation-Simulator (Op-Sim) \footnote{\url{https://github.com/lsst/rubin_sim}}using the Metrics Analysis Framework (MAF; \citealt{MAF}) \footnote{\url{https://www.lsst.org/scientists/simulations/maf}} \footnote{\url{https://github.com/lsst/rubin_sim_notebooks/tree/main}}.
\begin{figure*}
    \centering
    \includegraphics[width=1\linewidth]{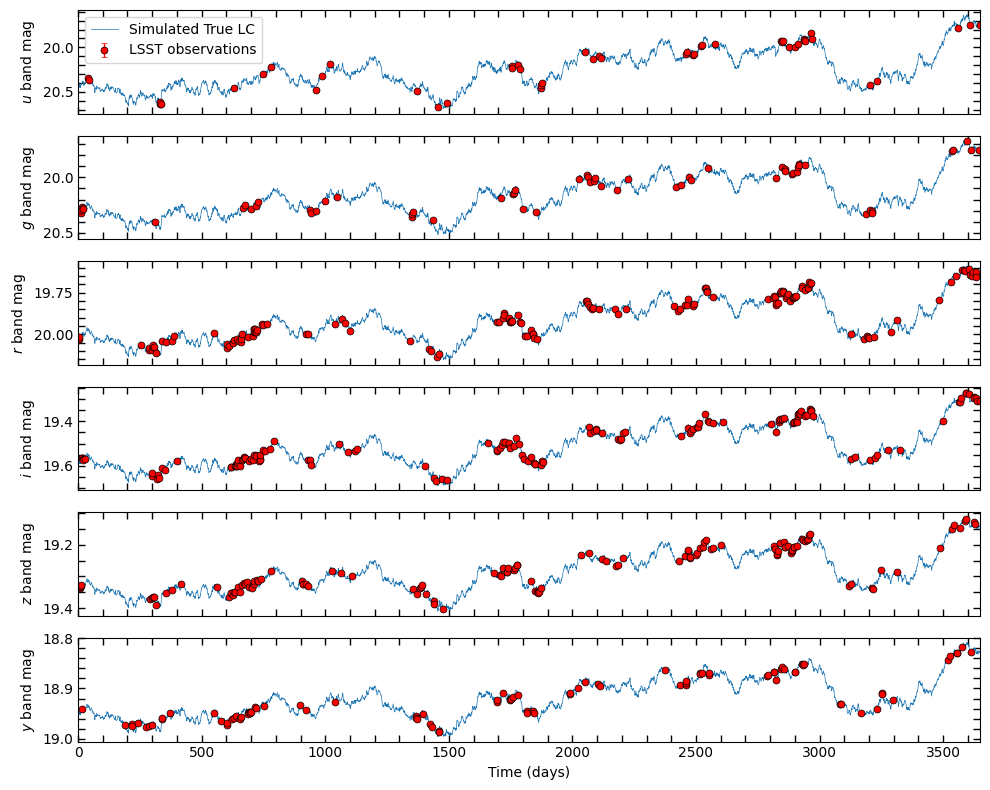}
    \caption{Simulated LSST light curve for a source using DRW parameters fit (SDSS J000008.13+001634.6	, $z$=1.836), as observed under the LSST Wide-Fast-Deep (WFD) survey strategy over the 10-year baseline. The underlying blue curve represents the light curve sampled with a uniform 1-day cadence. Overlaid red points correspond to LSST observations in all six bands ($ugrizy$), incorporating realistic survey cadence and photometric uncertainties based on the LSST observing strategy.}
    \label{fig:LSST_WFD_Cadence}
\end{figure*}

\begin{figure}[h]
    \centering
    \includegraphics[width=1\linewidth]{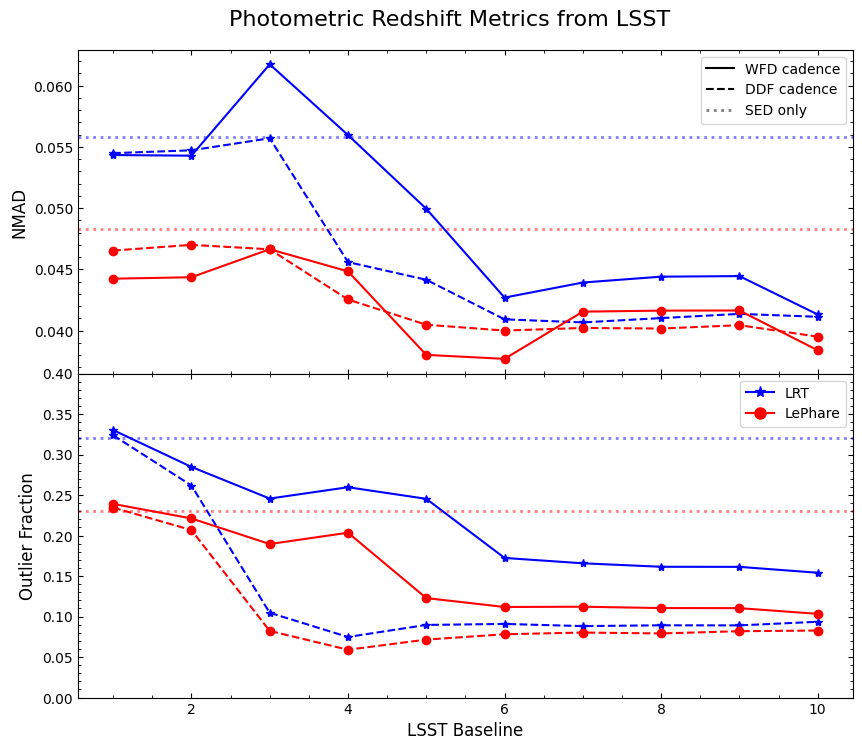}
    \caption{Evolution of photo-$z$ precision and outlier fraction for our parent sample as a function of the LSST temporal baseline. The panels also include a comparison with the results obtained using the LSST DDF cadence. This figure illustrates the improvements achieved by incorporating the \texttt{VAR-PZ} variability priors into both the \texttt{LRT} and \texttt{LePHARE} SED fitting routines, highlighting the impact on two independent methods. The \texttt{LePHARE} SED fitting is performed using the Sloan \textit{ugriz} filters, whereas both \texttt{LRT} and \texttt{VAR-PZ} utilize synthetic LSST \textit{ugrizy} filters.}
    \label{fig:-LSST_Metrics}
\end{figure}

Light curves were generated as a function of the LSST yearly data release. 
Figure \ref{fig:LSST_WFD_Cadence} illustrates the simulated 10-year DRW light curve for an AGN (SDSS J000008.13+001634.6), in the Wide-Fast Deep (WFD) \footnote{\url{https://survey-strategy.lsst.io/baseline/wfd.html}} survey. The improvement in the photo-$z$ performance of \texttt{LRT} obtained by adding the \texttt{VAR-PZ} priors are presented in Figure~\ref{fig:-LSST_Metrics}. As expected, the photo-$z$ precision improves and the outlier fraction decreases significantly over time, with the most limited performance seen during the first year. The figure also includes a comparison with photo-$z$ estimates derived using \texttt{LePHARE} SED templates, providing an independent validation. Note that, to avoid inconsistencies, the SED PDFs are obtained by fitting the Sloan \textit{ugriz} filters, instead of the simulated LSST stacked photometry (which relies on the \texttt{LRT} fits). Importantly, both \texttt{LePHARE} and \texttt{LRT} based routines, incorporating \texttt{VAR-PZ} variability priors, show consistent trends of improvement with increasing temporal baseline, further confirming the robustness of variability-informed photo-$z$ estimation. To assess the performance of \texttt{VAR-PZ} for the fainter AGNs that LSST will discover, we constructed a faint sample by dimming our original sources by a factor of three in flux. This setup approximates the parameter space where LSST will detect sources that are about three times fainter than those in SDSS, but with comparable $SNR$ owing to Rubin observatory's sensitivity and depth. A large fraction of faint AGNs detected by LSST will be host-dominated, for which SED-based photo-$z$s are expected to be relatively accurate even without variability information. In this experiment, however, we restrict ourselves to faint quasars.
For this faint sample, we simulated LSST light curves at the fainter magnitudes while keeping $R_{\mathrm{AGN}}$ fixed to the values inferred for the original sample.  We then combined the resulting \texttt{VAR-PZ} priors with the SED PDFs. For the faint quasars, our method reduces the outlier fraction from 32\% (SED alone) to 17\%,  with an NMAD of 0.04 when computed excluding outliers.

Additionally, we also test the level of improvement that would be obtained if these objects were instead within the Deep Drilling Fields (DDFs). Specifically, we assume the observing cadence of the from the XMM-Large Scale Structure Field (XMM-LSS\footnote{\url{https://survey-strategy.lsst.io/baseline/ddf.html}}; $\alpha_{\text{J2000}} = \SI{35.57}{\degree}$, $\delta_{\text{J2000}} = \SI{-4.82}{\degree}$). 
The DDF cadence yields improved photo-$z$ performance, with reduced outlier fractions and improved precision compared to the WFD strategy. Nevertheless, the WFD cadence still provides reasonably accurate redshift estimates for AGN populations, even though its typical outlier rate ($\sim15\%$) does not reach the higher fidelity achieved with the DDF fields ($\sim9\%$).

\section{Summary}\label{Summary}
In this work, we demonstrate that AGN variability can be effectively used to improve photometric redshift estimates. Specifically, we implement a method we refer to as \texttt{VAR-PZ} to generate redshift PDFs that can be applied as priors to, for example, photo-$z$ estimates based on SED modeling, serving not as a stand-alone solution for accurate AGN photo-$z$s but as a complementary tool to existing techniques.

This work serves as a proof of concept, motivating the development of other physically motivated variability models, which would likely enable tighter redshift constraints. 
As detailed in Section \ref{Methodology}, we model the observed stochastic variability of AGNs as a DRW process following the dependence of the DRW parameters on the physical properties of the AGNs proposed by M10. The parameters from M10 relation (Eq.~\eqref{equation}) are refitted to account for differences in light curve modeling and the AGN continuum luminosities, specifically $M_{\mathrm{i}}$ which is derived from SED modeling, along with host dilution fractions ($R_{\mathrm{AGN}}$).  While the DRW is a reasonable model for quasar variability, it should be emphasized that it is only an approximate description. Regardless, we show that it can yield broad but informative photo-$z$ PDFs.
 
To generate variability priors from \texttt{VAR-PZ}, we first estimate the values of the expected DRW parameters in each band as a function of redshift based on Eq.~\eqref{equation} in a redshift grid, performing a full SED decomposition to obtain $M_i$ and $R_{\mathrm{AGN}}$, following the methodology outlined in Section~\ref{Code}. We then compute the likelihood of those parameters being able to describe the observed light curves, and express the likelihood as a function of redshift as a PDF. We refer to this PDF as the \texttt{VAR-PZ} PDF.

To assess the effect of survey cadences and baselines, we took all AGN with well sampled light curves in the SDSS S82 (see Section \ref{data}) and we simulated SDSS-like light curves across a range of cadences and baselines (see Section \ref{Application}). Figure \ref{heatmap}, shows that the performance of photo-$z$s improve with longer and denser sampling. The results of SDSS observations (Section \ref{SDSS_real}), presented in Figure \ref{fig:zp_vs_zs_real} and Table \ref{tab:photz_performance}, demonstrate that the inclusion of \texttt{VAR-PZ} priors significantly improves redshift estimation. In comparison to the \texttt{LRT} SED modeling method by itself, the outlier fraction decreases from 27\% to 19\% for the full sample, and from 81\% to 33\% for sources with $z_{\mathrm{SED}} < 0.6$. When implementing \texttt{VAR-PZ} within the \texttt{LePHARE} SED–fitting framework, the performance is also improved albeit by a smaller factor, with outlier fractions reduced from 19\% to 16\% for the full sample and from 66\% to 37\% for low-redshift solutions. At higher redshifts, the limitations of SDSS observations, particularly the relatively short temporal baselines, reduce the reliability of variability–based redshift estimates (Figure \ref{zs_vs_tau}). In these regimes, the DRW timescales fall below one–third of the available light curve baseline, leaving the variability prior essentially unconstrained (See \citet{Suberlak_2021}, Section 2.3). This highlights that the gain from variability priors strongly depends on light curve baseline. Complementary simulations using Rubin LSST observing strategies (WFD and DDF) further confirm that LSST's increased temporal coverage and cadence substantially enhance photometric redshift accuracy. 
These findings, as shown in Figure ~\ref{fig:-LSST_Metrics}, demonstrate a reduction in the outlier fraction from $\sim$32\% for SED-only estimates to $\sim$15\% for WFD and $\sim$9\% for DDF cadences by the end of the survey. Additionally, for faint LSST objects, the WFD cadence yields an outlier fraction of $\sim17\%$ in separate tests. 
This underscores the importance of survey strategy and establishes the significance of incorporating variability-based priors into hybrid photometric redshift frameworks for upcoming large-scale AGN surveys.

\begin{acknowledgements}
We thank the anonymous referee for the thoughtful
comments that helped us improve the paper. We gratefully acknowledge the support of the ANID BASAL project FB210003 (S.S.S., R.J.A., T.A., F.E.B., C.M., T.M. and C.R.), FONDECYT Regular 1231718 (S.S.S. and R.J.A.), 1240105 (T.A.), 1241005 (F.E.B.), 1230345 (C.R) and ANID Millennium Science Initiative AIM23-0001 (T.A., F.E.B.). S.S.S. acknowledges the travel support from the Royal Astronomical Society to attend the meeting `Supermassive Black Hole studies in the Legacy Survey of Space and Time' in Durham, UK. T.T.A acknowledges support from NASA ADAP Grant 80NSSC24K0692. A.B. acknowledges support from the Australian Research Council (ARC) Centre of Excellence for Gravitational Wave Discovery (OzGrav), through project number CE230100016. DD acknowledges PON R\& I 2021, CUP E65F21002880003, and Fondi di Ricerca di Ateneo (FRA), linea C, progetto TORNADO. A.B.K. and D.I. acknowledge funding provided by the University of Belgrade - Faculty of Mathematics (the contract 451-03-136/2025-03/200104) through the grants by the Ministry of Science, Technological Development and Innovation of the Republic of Serbia. C.M. acknowledges support from Fondecyt Iniciacion grant 11240336. C.G.B. acknowledges support from the Consejo Nacional de Investigaciones Científicas y Técnicas (CONICET) and the Secretaría de Ciencia y Tecnología de la Universidad Nacional de Córdoba (SeCyT). C.M. acknowledges support from FONDECYT Iniciacion grant 11240336. D.M. acknowledges financial support from CAPES – Finance Code 001. D.D. and M.F. acknowledge the financial contribution from PRIN-MIUR 2022 and from the Timedomes grant within the "INAF 2023 Finanziamento della Ricerca Fondamentale". W.N.B. acknowledges the support of USA NSF grant AST-2407089. S.E.I.B. is supported by the Deutsche Forschungsgemeinschaft (DFG) under Emmy Noether grant number BO 5771/1-1. S.P. is supported by the international Gemini Observatory, a program of NSF NOIRLab, which is managed by the Association of Universities for Research in Astronomy (AURA) under a cooperative agreement with the U.S. National Science Foundation, on behalf of the Gemini partnership of Argentina, Brazil, Canada, Chile, the Republic of Korea, and the United States of America. C.R. acknowledges support from SNSF Consolidator grant F01$-$13252 and the China-Chile joint research fund. M.J.T. acknowledges funding from UKRI grant ST/X001075/1.

\newline
Funding for the SDSS and SDSS-II has been provided by the Alfred P. Sloan Foundation, the Participating Institutions, the National Science Foundation, the U.S. Department of Energy, the National Aeronautics and Space Administration, the Japanese Monbukagakusho, the Max Planck Society, and the Higher Education Funding Council for England. The SDSS Web site is \url{http://www.sdss.org/}.

The SDSS is managed by the Astrophysical Research Consortium for the Participating Institutions. The Participating Institutions are the American Museum of Natural History, Astrophysical Institute Potsdam, University of Basel, University of Cambridge, Case Western Reserve University, University of Chicago, Drexel University, Fermilab, the Institute for Advanced Study, the Japan Participation Group, Johns Hopkins University, the Joint Institute for Nuclear Astrophysics, the Kavli Institute for Particle Astrophysics and Cosmology, the Korean Scientist Group, the Chinese Academy of Sciences (LAMOST), Los Alamos National Laboratory, the Max-Planck-Institute for Astronomy (MPIA), the Max-Planck-Institute for Astrophysics (MPA), New Mexico State University, Ohio State University, University of Pittsburgh, University of Portsmouth, Princeton University, the United States Naval Observatory, and the University of Washington. 
\end{acknowledgements}

 \textit{Software packages}: \texttt{NumPy} \citep{2020Numpy}, \texttt{Astropy} \citep{2018Astropy}, \texttt{Matplotlib} \citep{Matplotlib}, \texttt{Celerite} \citep{Foreman-Mackey2017}, \texttt{Pandas} \citep{reback2020pandas}.

\bibliographystyle{aa}

\section*{Affiliations}
$^{1}$Instituto de Astrofísica, Facultad de Ciencias Exactas, Universidad Andres Bello, Fernández Concha 700, 7591538 Las Condes, Santiago, Chile \\
$^{2}$Instituto de Estudios Astrofísicos, Facultad de Ingeniería y Ciencias, Universidad Diego Portales, Av. Ejército Libertador 441, Santiago, Chile\\
$^{3}$Millennium Institute of Astrophysics, Nuncio Monseñor Sótero Sanz 100, Providencia, Santiago, Chile\\
$^{4}$European Southern Observatory, Karl-Schwarzschild-Strasse 2, 85748 Garching bei München, Germany\\
$^{5}$Max-Planck-Institut f\"ur extraterrestrische Physik, Giessenbachstr. 1, 85748 Garching, Germany\\
$^{6}$Department of Physics and Astronomy, Wayne State University, 666 W. Hancock St, Detroit, MI, 48201, USA\\
$^{7}$Instituto de Alta Investigación, Universidad de Tarapacá, Casilla 7D, Arica, Chile\\
$^{8}$School of Physics and Astronomy, Monash University, Clayton, Victoria 3800, Australia\\
$^{9}$ARC Centre of Excellence for Gravitational Wave Discovery -- OzGrav, Australia\\
$^{10}$Instituto de Astronomía Teórica y Experimental, (IATE, CONICET-UNC), Córdoba, Argentina\\
$^{11}$Universidad Nacional de Córdoba, Observatorio Astronómico de Córdoba\\
$^{12}$Institute for Theoretical Physics, Heidelberg University, Philosophenweg 12, D–69120, Heidelberg, Germany\\
$^{13}$Max-Planck-Institut f\"{u}r Astronomie, K\"{o}nigstuhl 17, 69117 Heidelberg, Germany\\
$^{14}$Department of Astronomy \& Astrophysics, The Pennsylvania State University, University Park, PA 16802, USA\\
$^{15}$The Institute for Gravitation for and the Cosmos, The Pennsylvania State University, University Park, PA 16802, USA\\
$^{16}$Department of Physics, 104 Davey Laboratory, The Pennsylvania State University, University Park, PA 16802, USA\\
$^{17}$Department of Physics, University of Napoli "Federico II", via Cinthia 9, 80126 Napoli, Italy\\
$^{18}$INAF - Osservatorio Astronomico di Capodimonte, via Moiariello 16, 80131 Napoli, Italy\\
$^{19}$Center for Theoretical Physics, Polish Academy of Sciences, Al. Lotników 32/46, 02-668 Warsaw, Poland\\
$^{20}$Dipartimento di Fisica “Ettore Pancini”, Università di Napoli Federico II, Via Cintia 80126, Naples, Italy\\
$^{21}$Ruđer Bošković Institute, Bijenička Cesta 54, 10000 Zagreb, Croatia\\
$^{22}$Frontier Research Institute for Interdisciplinary Sciences, Tohoku University, Sendai 980-8578, Japan \\
$^{23}$Global Center for Science and Engineering, Faculty of Science and Engineering, Waseda University, 3-4-1, Okubo, Shinjuku, Tokyo 169-8555, Japan\\
$^{24}$Department of Astronomy, Faculty of Mathematics, University of Belgrade, Studentski trg 16, 11000 Belgrade,
    Serbia\\
$^{25}$Hamburger Sternwarte, Universitat Hamburg, Gojenbergsweg 112, D-21029 Hamburg, Germany\\
$^{26}$Kavli Institute for Astronomy and Astrophysics, Peking University, Beijing 100871, People's Republic of China\\
$^{27}$Departamento de F\'isica, Universidad T\'ecnica Federico Santa Mar\'ia, Vicu\~{n}a Mackenna 3939, San Joaqu\'in, Santiago, Chile \\
$^{28}$Universidade Federal de Santa Maria (UFSM), Centro de Ciências Naturais e Exatas (CCNE), Santa Maria, 97105-900, RS, Brazil\\
$^{29}$International Gemini Observatory/NSF NOIRLab, Casilla 603, La Serena, Chile \\
$^{30}${Eureka Scientific, 2452 Delmer Street, Suite 100, Oakland, CA 94602-3017, USA}\\
$^{31}${Department of Physics, Yale University, P.O. Box 208120, New Haven, CT 06520, USA}\\
$^{32}$ NASA Goddard Space Flight Center, Greenbelt, MD 20771, USA \\
$^{33}$Center for Space Science and Technology, University of Maryland Baltimore County, MD 21250, USA \\
$^{34}$ Department of Astronomy, University of Geneva, ch. d’Ecogia 16,
1290, Versoix, Switzerland\\
$^{35}$Department of Physics, Drexel University, 32 S. 32nd Street, Philadelphia, PA 19104, USA\\
$^{36}$Exzellenzcluster ORIGINS, Boltzmannstr. 2, D-85748 Garching, Germany \\
$^{37}$Centre for Extragalactic Astronomy, Department of Physics, Durham University, South Road, Durham DH1 3LE, United Kingdom\\
$^{38}$Physics Department, Tor Vergata University of Rome, Via della Ricerca Scientifica 1, 00133 Rome, Italy\\
$^{39}$INAF – Astronomical Observatory of Rome, Via Frascati 33, 00040 Monte Porzio Catone, Italy\\
$^{40}$INFN – Rome Tor Vergata, Via della Ricerca Scientifica 1, 00133 Rome, Italy\\
$^{41}$Department of Physics \& Astronomy, Bishop’s University, 2600 rue College, Sherbrooke, QC, J1M 1Z7, Canada\\
$^{42}$National Radio Astronomy Observatory, 520 Edgemont Road, Charlottesville, VA 22904, USA\\
$^{43}$Department of Astronomy, University of Virginia, 530 McCormick Rd, Charlottesville, VA 22904, USA\\
$^{44}$Department of Astronomy, University of Michigan, 1085 S University, Ann Arbor, MI 48109, USA\\

\appendix
\FloatBarrier
\onecolumn
\section{VAR-PZ workflow}
\begin{figure}[h]
    \centering
    \includegraphics[width=\textwidth]{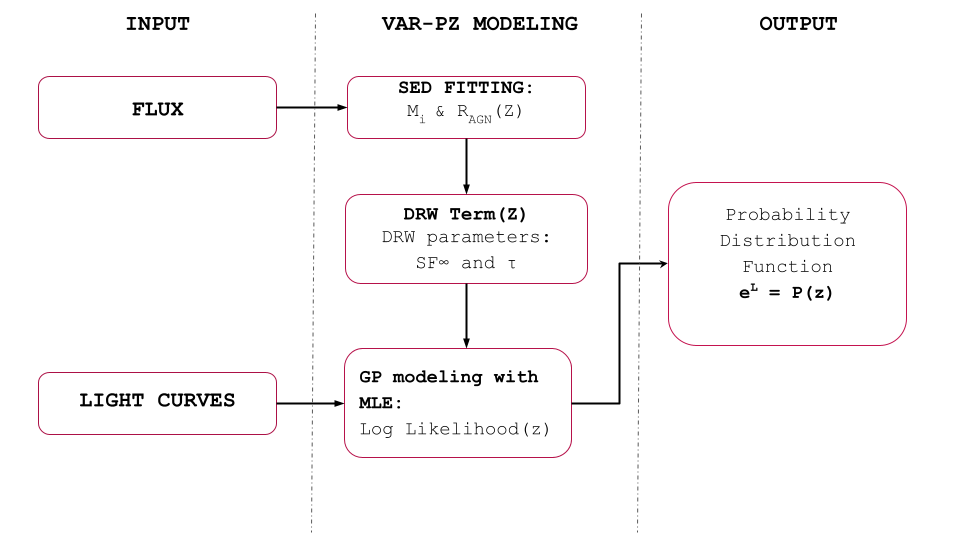}
    \caption{Flowchart illustrating the overall workflow of \texttt{VAR-PZ}: Input fluxes and light curves are used to model the SED using \texttt{LRT} templates, and the light curves are modeled as a DRW using the M10 kernel. The resulting kernel is incorporated into a Gaussian Process to produce redshift-dependent photometric redshift PDFs.}
    \label{fig:routine_flowchart}
\end{figure}

Figure \ref{fig:routine_flowchart} presents the computational pipeline and algorithmic steps used in the \texttt{VAR-PZ} implementation.
This flowchart serves as a visual reference for the complete \texttt{VAR-PZ} computational workflow.

\twocolumn

\onecolumn

\section{LSST DDF cadence}
\begin{figure}[h]
    \centering
    \includegraphics[width=\textwidth]{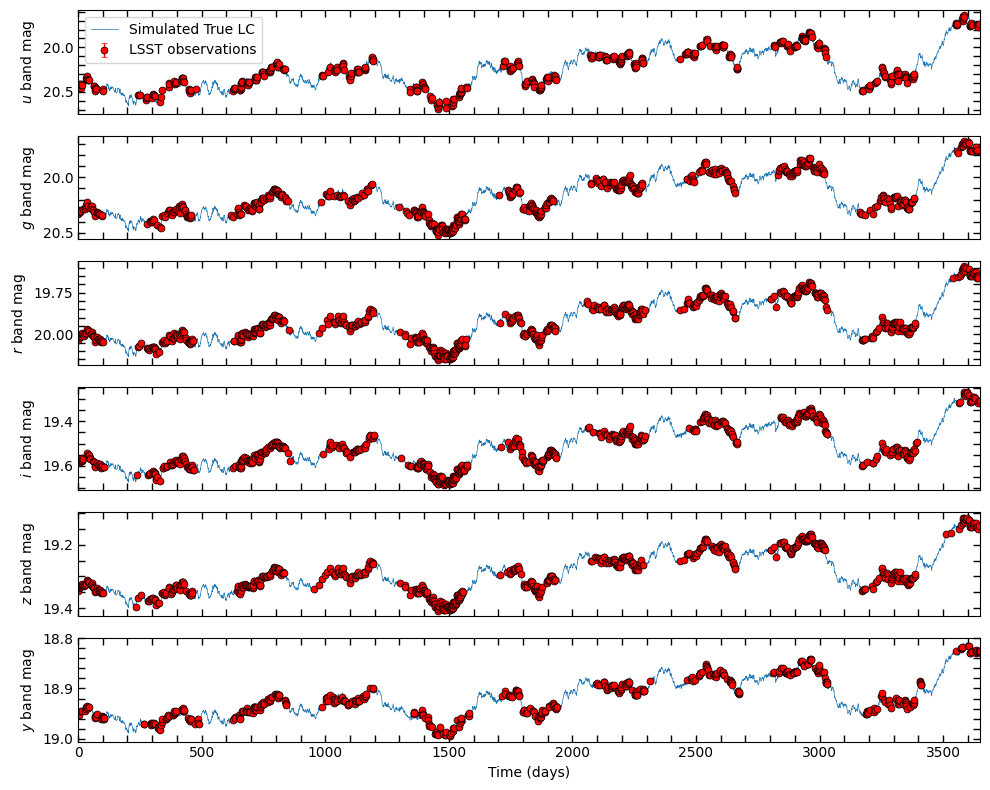}
    \caption{Simulated light curve for an AGN located in a Rubin LSST DDF. This example corresponds to the XMM-LSS field, centered at $\alpha_{\text{J2000}} = \SI{35.57}{\degree}$, $\delta_{\text{J2000}} = \SI{-4.82}{\degree}$}
    \label{fig:lsst_cadence_ddf}
\end{figure}

Figure \ref{fig:lsst_cadence_ddf} a simulated light curve based on the XMM-LSS DDF observing cadence to illustrate the improved temporal sampling compared to the WFD.
This simulation highlights the dense sampling of the DDFs, which constitute approximately 6.5\% of the total survey time and are designed to provide intensive monitoring, with each field receiving over 20,000 visits over the survey duration.

\end{document}